\documentclass[12pt]{article}
\usepackage{amsmath}
\usepackage{amssymb}
\usepackage[top=1.25in, bottom=1.25in, left=1.25in, right=1.25in]{geometry}

\usepackage{fancyhdr}

\begin{document}
\title{New Setting for Spontaneous Gauge Symmetry Breaking?}
\author{R. Jackiw$^*$ and S.Y. Pi$^{\dagger \S}$\\
\textit{\small $^*$CTP, Massachusetts Institute of Technology}\\
\textit{\small $^\dagger$Department of Physics, Boston University} 
}
\date{}

\maketitle
\thispagestyle{fancy}

\begin{abstract}
Over half century ago Carl Brans participated in the construction of a viable deformation of the Einstein gravity theory. Their suggestion involves expanding the tensor-based theory by a scalar field. But  experimental support has not materialized. Nevertheless the model continues to generate interest  and new research. The reasons for the current activity is described in this essay, which is dedicated to Carl Brans on his eightieth birthday. 

$^\S$This research was supported in part by U.S. Department of Energy, Grant No. DE-SC0010025.
\end{abstract}

C. Brans and R. H. Dicke (also P. Jordan) proposed a tensor/scalar ($g_{\mu\nu}/\varphi$) generalization of Einstein's general relativistic tensor gravity model.  In their generalization a scalar field $\varphi$ is coupled to the Ricci scalar, and further dynamics is posited for $\varphi$. The dynamical equations follow from a generalized Einstein-Hilbert Lagrangian
\[
\begin{array}{lll}
I_\alpha &= & - \int \mathcal{L}_\alpha\\[1ex]
\mathcal{L}_\alpha &=&   \sqrt{-g} \left[\frac{\alpha}{12} \ \varphi^2 R + \frac{1}{2}\ g^{\mu\nu} \, \partial_\mu \varphi  \, \partial_\nu \varphi+ \lambda \varphi^4 \right]
\end{array}
\]
The parameter $\alpha$ measures the strength of the $R-\varphi^2$ interaction and suggests a dynamical origin for the gravitational constant $G \propto 1/\ \varphi^2$. (A self coupling of strength $\lambda$ may also be included, but it plays no role in our present discussion.)\cite{jPhysRev.124.925}

While the model is attractive in that it presents a very explicit modification of the  Einstein theory, it fails to agree with the experimental values for the classic solar system tests of gravity theory. Nevertheless these days interest has revived in the  Brans-Dicke model at $\alpha=1:  \mathcal{L}_W=  \mathcal{L}_{\alpha = 1}$.
\begin{equation}
 \mathcal{L}_W =\sqrt{-g} \left\{\frac{1}{12} \ \varphi^2 R + \frac{1}{2}\ g^{\mu\nu} \, \partial_\mu \varphi\, \partial_\nu \varphi + \lambda \varphi^4\right\}
\label{eq1}
\end{equation}
The $\alpha= 1$ model possesses Weyl invariance, \textit{i.e} invariance against rescaling the dynamical variables by a local space-time transformation.
\begin{subequations}\label{eq2}
\begin{eqnarray}
g^{\mu\nu} & \to & e^{2\theta} g^{\mu\nu}\label{eq2a}\\[1ex]
\varphi & \to & e^\theta \varphi\label{eq2b}
\end{eqnarray}
\end{subequations}
Here $\theta$  is an arbitrary function on space time. The reasons for the contemporary interest in $ \mathcal{L}_W$ are the following.

These days physicists are satisfied by the success that has been achieved in understanding and unifying all forces save gravity. This has been accomplished with the help of spontaneous breaking of local internal symmetries. 

With the desire to include gravity in this framework, and in keeping with its presumed geometric nature, various people have suggested  studying Weyl invariant dynamics, with the hope that Weyl scaling will help understand  short distance phenomena. Additionally some   are tantalized by the long-standing desire to extend conventional space-time symmetries to include local conformal (Weyl) symmetry \cite{Hooft:2014daa}.  

$ \mathcal{L}_W$ seems to bring closer  the above goals: An operative Weyl symmetry appears to host a local gauge symmetry, which can be broken by choosing specific values for $\varphi$. Indeed $\varphi =1$ renders $ \mathcal{L}_W$ equal to the Einstein-Hilbert Lagrangian. In this framework Einstein theory is merely the ``unitary gauge" version of $ \mathcal{L}_W$.

Certainly such ideas are provocative and worthy of further examination and possible development. However, a critical viewpoint leads to the following questions and observations.
\begin{enumerate}\setlength{\itemsep}{-.5ex}
\item No gauge potential (connection) is present;  in what sense does $I_W$ define a ``gauge theory" ?
\item There is no dynamical/energetic reason for choosing the ``unitary gauge" $\varphi = 1$. (In familiar spontaneous breaking, asymmetric solutions are selected by lowest energy considerations.)
\item By inverting the order of presentation, we recognize that $\varphi$   is a spurion  variable: upon replacing  $g_{\mu\nu}$      in the Einstein-Hilbert action by $g_{\mu\nu} \varphi^2$  , one arrives at the Weyl action \cite{PhysRevD.3.1689}.
\[
I_{\text{Einstein-Hilbert}}\,  \big|_{g_{\mu\nu} \to\, g_{\mu\nu}\,  \varphi^2 }  \to I_W 
\]
\item The Weyl symmetry current vanishes identically. The computation is performed  according to Noether's first theorem (applicable when transformation parameters are constant) and her second theorem (applicable when transformation parameters depend on space-time coordinates). The former is a special case of the latter; both give the same result: no current. With no current, there is no charge and no symmetry generator.
\end{enumerate}

The fact that the Weyl current vanishes cannot be attributed to the locality of the symmetry transformation parameter $\theta(x)$. An instructive example is electrodynamics, where $\delta A_{\mu} =\partial_{\mu} \theta$ and $\delta\Psi =- i \theta \Psi$ for a charged field $\Psi$. The current is non-vanishing and is identically conserved, i.e. it is a superpotential.
\begin{equation}
J^\mu = \partial_\nu\, (F^{\mu\nu}\, \theta)
\label{jackso16}
\end{equation}
(This is the Noether current for gauge symmetry, not the source current $J^\mu_{EM}$  that appears in the Maxwell equations.) While the dependence on an inhomogenous $\theta$ may make $J^\mu$ unphysical, the global limit produces a sensible result. 
\begin{equation}
J^\mu = \partial_\nu\, (F^{\mu\nu})\, \theta = J^\mu_{EM} \theta
\label{jackso17}
\end{equation}
In the Weyl case, setting the parameter $\theta$ to a constant leaves a global symmetry. Yet the current still vanishes.

Evidently the vanishing of the Weyl symmetry current, both local and  \textit{a forteriori} also global, reflects the particularly peculiar role of the Weyl ``symmetry" in the examined models \cite{PhysRevD.91.667501}.

As yet we do not know how to assess the significance of the above observations for a physics program based on  Weyl symmetry. Clearly it is interesting to explore the similarities to and differences from the analogous structures in conventional gauge theory. We conclude with two observations on the model.

 By  an alternate ``gauge choice" we can set $\sqrt{-g}$ to a constant. Evidently, a unimodular scalar/tensor theory is ``gauge equivalent" to the Einstein-Hilbert model.

The kinetic term for $\varphi$ is not Weyl invariant and its non-invariance  is compensated by  the non-minimal  interaction with $R$. Alternatively we may dispense with the non-minimal interaction and achieve  invariance by introducing a gauge field $W_\mu$, which transform as \mbox{$W_\mu \to  W_\mu - \partial_\mu \theta$}. One verifies invariance of 
\begin{alignat}{3}
\int d^4 x \, \sqrt{-g} \ \bigg(\frac{1}{2}\ g^{\mu\nu}\, & D_\mu  \varphi\   D_\nu\,  \varphi  \bigg)\nonumber\\ 
& D_\mu \equiv \partial_\mu + W_\mu . &
\label{jackEq5}
\end{alignat}
Expanding and integrating by  parts shows that \eqref{jackEq5} is equivalent to \eqref{eq1} provided $R$ is given by the formula  \cite{Iorio1997}
\begin{equation}
\frac{1}{12} \ R = D^\mu W_\mu + \frac{g^{\mu\nu}}{2}\, W_\mu W_\nu .
\end{equation}

\noindent While these observations are provocative, they have not produced any useful insights. Indeed thus far the only  established role for $I_W$ is to   generate  the traceless new improved energy momentum tensor $\theta^{CCJ}_{\mu\nu}$ \cite{Callan1970}:
\begin{equation}
\theta^{CCJ}_{\mu\nu} = \frac{2}{\sqrt{-g}}\ \frac{\delta I_W}{\delta g^{\mu\nu}}\  \bigg|_{g_{\mu\nu} \to\, \delta_{\mu\nu}}
\end{equation}

\end{document}